\documentclass{PoS}
\usepackage{polski}
\usepackage[utf8]{inputenc}
\usepackage{graphicx}
\usepackage{enumerate}
\usepackage{caption}
\usepackage{xcolor,colortbl}

\definecolor{LightRed}{rgb}{238, 210, 210}
\definecolor{LightGrey}{rgb}{208,208,208}

\newcolumntype{a}{>{\columncolor{LightRed}}c}
\newcolumntype{b}{>{\columncolor{LightGrey}}c}

\title{A communication solution for portable detectors of the ``Cosmic Ray Extremely Distributed Observatory''}

\ShortTitle{Communication solution for CREDO portable detectors }

\author{\speaker{Katarzyna Smelcerz$^{1}$}, Konrad Kopański$^{2}$, Wojciech Noga$^{3}$, Mateusz Sułek$^{2,3}$, Kevin Almeida Cheminant$^{2}$, Niraj Dhital$^{2}$, Dariusz Gora$^{2}$, Piotr Homola$^{2}$, Oleksandr Sushchov $^{2}$, Dmitriy Beznosko$^{4}$, Jilberto Zamora-Saa$^{5}$, Alan R. Duffy$^{6}$, Marcin Kasztelan$^{7}$, Peter Kovacs$^{8}$, Vahab Nazari$^{2,9}$, Michał Niedźwiecki$^{1}$, Krzysztof Rzecki$^{3}$, Karel Smolek$^{10}$, Jaroslaw Stasielak$^{2}$, Zoltán Zimborás$^{11}$ \\
         E-mail: \email{
        	katarzyna.smelcerz@pk.edu.pl\\
        }
        $^{1}$Institute of Telecomputing, Faculty of Physics, Mathematics and Computer Science, Cracow University of Technology, Warszawska 24st 31-155 Cracow, Poland\\
        $^{2}$Institute of Nuclear Physics Polish Academy of Sciences, Radzikowskiego 152, Cracow, Poland\\
        $^{3}$Cracow University of Technology, Warszawska 24, 31-155 Cracow, Poland\\
        $^{4}$Bard College, New Orleans Louisiana, United States\\
        $^{5}$Universidad Andres Bello, Departamento de Ciencias Fisicas, Facultad de Ciencias Exactas, Santiago, Chile\\
        $^{6}$Centre for Astrophysics and Supercomputing, Swinburne University of Technology, Hawthorn, VIC 3122, Australia\\
        $^{7}$National Centre for Nuclear Research, Andrzeja Soltana 7, 05-400 Otwock-Swierk, Poland\\
        $^{8}$Institute for Particle and Nuclear Physics,Wigner Research Centre for Physics, Hungarian Academy of Sciences, H-1525 Budapest, Hungary\\
        $^{9}$Joint Institute for Nuclear Research, Dubna, Russia\\
        $^{10}$Institute of Experimental and Applied Physics, Czech Technical University in Prague\\
        $^{11}$Wigner Research Centre for Physics of the Hungarian Academy of Sciences
        
       }

\abstract{The search for Cosmic-Ray Ensembles (CRE), groups of correlated cosmic rays that might be distributed over very large areas, even of the size of the planet, requires a globally spread and dense network of detectors, as proposed by the Cosmic-Ray Extremely Distributed Observatory (CREDO) Collaboration. This proposal motivates an effort towards exploring the potential of using even very much diversified detection technologies within one system, with detection units located even in hard-to-reach places, where, nevertheless, the sensors could work independently – without human intervention. For these reasons we have developed a dedicated communication solution enabling the connection of many different types of detectors, in a range of environments. The proposed data transmission system uses radio waves as an information carrier on the 169MHz frequency band in a contrast to the typical commercially used frequencies in a IoT systems (868MHz). The connectivity within the system is based on the star topology, which ensures the least energy consumption. The solution is now being prepared to being implemented using the prototype detection system based on the CosmicWatch open hardware design: a portable, pocket size, and economy particle detector using the scintillation technique. Our prototype detector is equipped with a dedicated software that integrates it with the already operational CREDO server system.}

\FullConference{36th International Cosmic Ray Conference -ICRC2019-\\
		July 24th - August 1st, 2019\\
		Madison, WI, U.S.A.}

\begin{document}

\section{Introduction}
Due to the study of Cosmic-Ray Ensembles (CRE) in large areas and the correlation between specific events, the project requires a large and widely dispersed research infrastructure. CREDO wants to rely not only on specialized detectors, such as those found in professional observatories (eg. Pierre Auger), which are available only to the scientific staff, but also on detectors that could be used by people who are not necessarily scientists (citizen science) [1]. Detectors that could be used in a large-scale experiment should be generally available, easy to use, and above all - cheap. This approach resulted in the implementation of smartphones in the CREDO project as cosmic ray detectors [2]. It is estimated that nowadays about 2.7 billion people are smartphone users [3]. Telephones as detectors unfortunately have quite limited "field of view" - the matrix has an average size of 1/2.3”, also require permanent connection to the charger while operating the CREDO application. Thus, in a scenario in which we would like to "catch" particles in areas that are difficult to access, i.e. forests or deserts, mobile phones will not pass the exam. Detectors according to the scenario are to be deployed in both urbanized and uninhabited areas. This means that not everywhere will be access to the Internet, 3G or GSM.

An alternative to the GSM or WiFi network is ISM bandwidth, which does not require any infrastructure. Available ISM bands are 915 MHz, 868 MHz, 433 MHz and 169 MHz. The largest range is obtained for the lowest frequencies, because the attenuation of the signal through obstacles is proportional to the square of the frequency [4] (Formula \ref{kwadrat}). The aim was to select a universal carrier that will provide communication in any environment. Therefore, the ISM 169 MHz band was selected for transmitting information in CREDO scenario.

\begin{equation} 
L = L_{x}f^2
\label{kwadrat}
\end{equation} 
where $L_{x}$ is constant characteristic to the material of the obstacle.

\section{System Topology}
The system is organized in the star topology, which ensures the lowest energy consumption (does not require the use of retransmitters). The center is a collecting station (sink) equipped with a specialized antenna for communication at the frequency of 169 MHz. The sink being in the mode of continuous listening, waiting for the packages with data that are sent by the broadcasting stations requires mains power. Sink is connected to the computer by a COM port through which it transmits data received from transmitting stations to the terminal.

\begin{figure}
	\centering
	\includegraphics[scale=0.5]{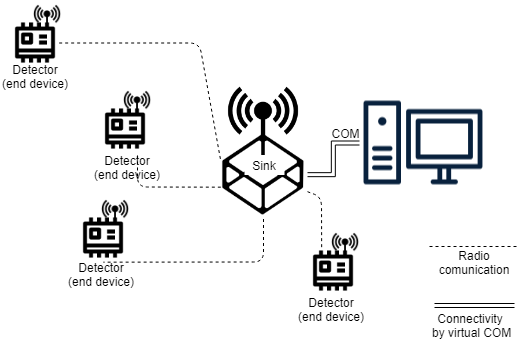}
	\caption{Topology of CREDO communication system}
	\label{star_topology}
	%\ref{star_topology}
\end{figure}

Transmission stations are intended for integration with mobile devices (in the case of the CREDO project with mobile detectors), therefore, when designing them, particular emphasis was placed on minimizing energy consumption. The diagram of the system architecture is shown in Figure \ref{star_topology}.

\noindent

\newpage
\vspace{5cm}

\section{Radio waves as a carrier of wireless communication}
There are many ready-made solutions on the market that provide wireless communication on ISM bands, eg. LoRa and SigFox [5]. Unfortunately, the use of such solutions requires infrastructure (receivers placed on masts), which are only available in larger cities. In this case, it is not possible to use the communication system in non-urbanized areas.
The range and mobility of the station is a critical parameter when choosing the technology for the communication system for the CREDO project. During the system design, tests were carried out on starter-kit STM equipped with LoRa modules [6]. However, the LoRa protocol was not used because it was known that it could not be used in non-urbanized locations and only the LoRa’s modulation mode was tested. Unfortunately, no satisfactory coverage has been achieved (~300 m in urban area). Similar tests were conducted on starter-kit sets equipped with module SPIRIT1 [7], similarly, satisfactory test results were not obtained - in the built-up area the range was 100 m.
Due to the fact that it is required to provide communication for data readings from detectors both in urban and secluded areas, it finally became clear that it would be necessary to design our own broadcasting and receiving stations that were autonomous and dedicated to the project.

\section{Transmission protocol}
\begin{figure}
	\centering
	\includegraphics{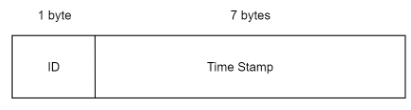}
	\caption{Frame format sent to the sink}
	\label{frame}
	%\ref{star_topology}
\end{figure}

The system is based on one-way transmission of detectors to the receiving station. To avoid collisions between individual transmissions, clock synchronization based on the DCF clock and transmission in the "time slots" designated by the ID number of the transmitting stations, was used. In places where the DCF clock is unavailable, the transmitting station has the ability to emulate the DCF signal. Each of the transmitting stations sends the frame to the sink in case of an event (particle detection) in the nearest "time slot". In addition to that, once an hour a status frame is sent, including its GPS position.
The DCF signal is transmitted on the 77.5 kHz frequency from Germany and can be received within a radius of 2000 km. In places where DCF reception is not possible, the receiving station emulates a DCF signal to synchronize the time in broadcasting stations.
Each of the transmitting stations is assigned an ID number, which is an integer in the range $<1,+\infty)$. In the experimental system, it was assumed that the transmitting stations within one receiving unit is at most 60 pcs. and that they have ID numbers from 1 to 60. These ID numbers correspond to time slots $n$, where $n = IDnumber -1$, i.e. numbers from 0 to 59 corresponding to subsequent minutes.
The format of the frame sent by the sending station is presented on the Figure \ref{frame}.

\noindent
The end frame marker is not necessary, as a used radio module Semtech [8] recognizes the end of the frame on its own.
The average life time of a battery at the sending station is presented below is ~2 years in a scenario where the data is sent once per hour from a detector, and the frame transmission time is 2.2 s.
The battery lifetime was calculated by Formula \ref{bateria}.

\begin{equation} 
battery life [h] =\frac{B [mAh]}{\bar{I} [mA]}
	\label{bateria}
\end{equation} 
where $B$ is battery capacity, and $\bar{I}$ is average current.

The average current was calculated by Formula \ref{srednia}.

\begin{equation} 
	\bar{I}[A] =\frac{t[s] p[A] n}{q[s]}
	\label{srednia}
\end{equation}

where $t$ is transmission time, $p$ is power consumption when transmitting, $n$ is number of frames per day and $q$ is the number of seconds per day.

\section{Antenna}
Matching the right type and parameters of the antenna had a huge impact on the results obtained during the experiment. The experiment showed that the type of antenna not only affects the range but also the quality of the transmission.
The basic parameters of the antenna are:
\begin{itemize}
	\item Gain in relation to the reference antenna (dipole with a length of $\frac{1}{2}$ wave)
	\item Impedance
	\item Directionality
\end{itemize}

Wavelength is another important parameter for determination of antenna length - the distance that the radio wave travels during one complete cycle of the wave.

The project uses antennas with $50\Omega$ impedance. The required length of the rod antenna should be at least $\frac{1}{4}$ of the radio wavelength. This means, the lower frequency, the longer antenna rod is required; it is a serious problem for mobile devices that should be miniaturised.
Small size constructions use antenna other than rods, which are shorter with good gain. The project uses transmitting antennas with a length of 110 mm and a gain of 2 dBi, while the rod antenna would have to have a length of about 400 mm.
The antenna in the receiving station does not have to be miniaturised. A full wave antenna (approx. 1500 mm long) and a gain of over 10 dBi was used.

\section{The scope of the coverage experiment}
\begin{figure}
	\centering
	\includegraphics{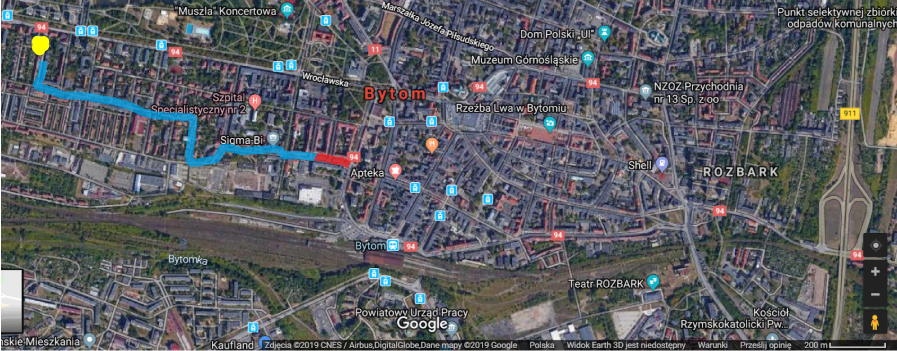}
	\caption{Satellite map, source: https://www.google.pl/maps}
	\label{path_map}
	%\ref{path_map}
\end{figure}

The experiment was conducted in the most difficult of possible conditions, including potential scenarios of using the communication system in the CREDO project - meaning a high urban development in the city of Bytom. It should be remembered that the waves are reflected and absorbed by buildings, hence the range in urban areas is always smaller than in open areas. Figure \ref{path_map} shows the satellite map of the area in which the trials were carried out. About 20 tests were carried out with various types of amplifiers and radio modules and antennas on the track marked in color.
The system consisted of one transmitting station, where the 8 bytes size frame was sent every 5 s, and one receiving station. The transmitting station (adapted to work with mobile devices) was mixed up at a walking pace. The person who carried out the experiment went for a walk with the sending station. Receiving station (sink) is marked with a yellow dot on the Figure \ref{path_map}.
The sending station is based on a standard implementation of the SX1276, transceiver LoRa [8]. The sink station is equipped with same type of the transceiver, however it has added very sensitive amplifier LNA.
Places where communication failed - frames from the transmitting station did not reach the sink at all, are marked in red on the Figure \ref{path_map}. The best results have been obtained so far for the configuration presented in the current work. 
The system configuration may seem very simple, but it should be noted that a more complicated configuration is not required to test the transmission range.

To carry out the communication coverage test, the transmitting station sends data about the transmission, which are then collected at the terminal on the computer. The data is the device $ID$, $frame  number$, $RSSI$ and $SNR$, where:

SNR (Signal Noise to Radio) is described by the Formula \ref{snr_long}:
\begin{equation} 
SNR(dB) =10\log_{10}\frac{P_{signal}}{P_{noise}}
\label{snr_long}
\end{equation} 

if all levels are expressed in decibels, then the formula can be simplified to the Formula \ref{snr_short}

\begin{equation} 
SNR(dB) = P_{signal}(dB) - P_{noise}(dB)
\label{snr_short}
\end{equation}
and RSSI (Received Signal Strength Indication), that is the value of the signal measured in decibels (dB) from $0$ to $-120$. The closer to $0$, the stronger the signal is.
Some examples of collected frames during the experiment are presented in Table \ref{tab}. Data marked on blue are form the parts of blue area from the Figure \ref{path_map}. Data marked on red are from parts of the red area from the Figure \ref{path_map}. The data which did not arrived to the sink are marked with $x$. 
The transmission parameters for the entire path shown on Figure \ref{path_map} can be found at https://github.com/kashiakashia/The-CREDO-communication-system/wiki
\vspace{0.5cm}

\begin{table}
	\centering

\begin{tabular}{| >{\columncolor[RGB]{230, 242, 255}}p{0.5cm} |>{\columncolor[RGB]{230, 242, 255}} p{0.7cm} |>{\columncolor[RGB]{230, 242, 255}} p{1cm} |>{\columncolor[RGB]{230, 242, 255}} p{1cm} |>{\columncolor[RGB]{230, 242, 255}} p{1cm}|||>{\columncolor[RGB]{238, 210, 210}} p{0.5cm} |>{\columncolor[RGB]{238, 210, 210}} p{0.7cm} |>{\columncolor[RGB]{238, 210, 210}} p{1cm} | >{\columncolor[RGB]{238, 210, 210}}p{1cm} |>{\columncolor[RGB]{238, 210, 210}} p{1cm} |}
	\hline
	n &	ID & frame & RSSI &	SNR & n&ID & frame & RSSI &	SNR	\\
	\hline
	\hline
	392&	 12&	597&  -49&  6&  1&  x&  160&  -69&  -16\\
	393&	 12&	598&  -55&  12&   2&  x&  165&  -70&  -15\\
	394&	 12&	599&  -55&  11&	  3&  12&	167&	-69& -15\\
	395&	12&	600&	-54&	11&	  4&	12&	168&	-69& -15\\
	
	396&	12&	601&	-50&	7&	5&	x&	169&	-69& -16\\
	
	397&	12&	602&	-47&	6&	6&	x&	x&	-69&	-17\\
	
	398&	12&	603&	-45&	9&	7&	x&	176&	-69&	-15\\
	
	399&	12&	604&	-44&	7&	8&	12&	x&	-69&	-16\\
	
	400&	12&	605&	-44&	7&	9&	x&	x&	-68&	-17\\

	401&	12&	606&	-44&	7&	10&	x& x&	-76&	-17\\
	
	402&	12&	607&	-44&	6&	11& x& x&	-69&	-16\\
	
	403&	12&	608&	-44&	7&	12&	12&	182&	-69&	-15\\
	
	404&	12&	609&	-55&	7&	13&	x&	x&	-77&	-17\\
	
	405&	12&	610&	-52&	12&	14&	12&	186&	-68&	-17\\
	
	406&	12&	611&	-53&	11&	15&	12&	4&	-68&	-17\\
	
	407&	12&	612&	-52&	11&	16& x&	189&	-69&	-15\\

	\hline

	%\caption{a}
	%\label{path_map}
\end{tabular}
  \caption{Transmission parameters from the range limit area (red) compared with transmission parameters for the very good range area (blue)} \label{tab}
\end{table}

\section{New $\mu C$ for CosmicWatch}
One of the devices, able to detect cosmic rays, is an MIT product called CosmicWatch (CW) [9]. CW is a small device contains Silicon Photomultiplier (SiMP) diode attached to the plastic scintillator and electronic for analyze and save incoming events. Heart of the CW is ATmega328P (ATmega) microprocessor ($\mu C$) working on 16 MHz, without another modules. During the test, CW have problems with time slew. After a few hours the two ATmega32s without synchronization can have different times which is a challenge for data analysis. A solution to this problem is to use a GPS module which for the STM32 device family worked acceptably.

As a test the STM32F446ZE (STM) on 180 MHz was used, with FGPMMOPA6H GPS module [10][11]. The internal two counters are used for timestamp for nanoseconds time frame and one timer for synchronization with GPS. In fact, the accuracy of time depends mainly from GPS module and conditions of GPS signal. Ideally this GPS module should have only 10 ns jitter, but in fact this parameter is worse. Replacing $\mu C$ in CW to STM with GPS module provide great increase of performance and time precision than original solution of CW. STM is 11.25 times faster than original used ATmega and direct memory access (DMA) used in STM is solution for dead time caused by communication with GPS module and data receiver (server) existing in ATmega. 

While better software or using bespoke hardware such as FPGA can further improve the system it is not required. Our aim was to achieve an optimal price to performance ratio with sufficient timing precision, for that the $\mu C$ solution was satisfactory. The software to connect CW and solution on STM with CREDO database is created software written in Python available at https://github.com/credo-science/Credo-Desktop-Detector.

\section{Conclusion and future work}

To obtain a fully autonomous and mobile system for the detection of CRE, it is necessary to create our own low-power and mobile detector equipped with a reference real time clock. Unfortunately, CosmicWatch detectors take too much electricity to power them with the battery, and synchronization of their time base is not satisfactory for very accurate measurements. An important aspect is also further improvement of the communication system. At the moment we are moving through 2D space; going down to lower frequencies, such as 27 MHz,  would allow the possibility of communication through the the ionospheric reflection of radio waves. This would allow the bypassing of obstacles (such as high buildings, ascents) in 3D space.

Assuming the detection station has access to a power supply, it would be worthwhile using simple systems to control the climatic conditions inside the measuring station. In the prototype station the Peltier Modules and the water cooling system were used, to estimate its efficiency and reliability. Low power heaters were also used to dry the station interior and guarantee a safe and convenient operating temperature, even in temperatures below zero (centigrade). The system is also equipped with condensate pumps and an emergency ventilation system, based on compressed air, inside the chamber. Taking into account the physics of the detection process, the key element is to ensure the appropriate conditions, mainly thermal ones. In addition, most plastic scintillators significantly change their properties after repeated exposure to temperature changes above certain limits. When using the detection technique based on semiconductor components (SiPM, amplifiers) in the analog path, the temperature differences can also significantly affect the measurement results. For this purpose, it is necessary to conduct accurate thermal tests of the detectors used in a station, and to select the optimal measurement conditions for them, using active methods (controlled, autonomous climate chambers), as well as passive ones - radiators and housings allowing the most favorable heat distribution. 

By optimizing detection techniques for use in a stations with the lowest possible power consumption (battery, photovoltaic), it is necessary to choose the proper detection technique for changeable environmental conditions. It is therefore advisable to make thermal calibration of analog circuits, and to take into account the results of calibration in subsequent data analysis. It is also necessary to test the influence of temperature on the scintillator parameters and to select the right material, guaranteeing the best results in a wide temperature range, while maintaining the lowest possible price and highest durability.


\begin{thebibliography}{99}
\bibitem{...}
“We are all the Cosmic-Ray Extremely Distributed Observatory”, N. Dhital, et al. (CREDO Collab.), PoS (ICRC2017) 1078 [arXiv:1709.05196]

\bibitem{...}
“Cosmic-Ray Extremely Distributed Observatory: status and perspectives”,D.Góra, et al. (CREDO Collab.), Universe 2018,4(11) 111 .[arXiv:1810.10410]

\bibitem{...}
Statistic data https://www.statista.com/statistics/330695/number-of-smartphone-users-worldwide/, available on 5th July 2019

\bibitem{...}
“Urban Propagation Modeling For Wireless Systems”, William Mark Smith, Donald C. Cox, A Report submitted to the Army Research Office in fulfilment of Short Term Innovative Research (STIR) Grant DAAD19-03-1-0069

\bibitem{...}
“An Evaluation Of Low Power Wide Area Network Technologies For The Internet Of Things”, Keith E. Nolan, Wael Guibene, Mark Y. Kelly, 2016 International Wireless Communications and Mobile Computing Conference (IWCMC) 

\bibitem{...}
Technical documentation for B-L072Z-LRWAN1 LoRa®/Sigfox™ Discovery kit https://www.st.com/en/evaluation-tools/b-l072z-lrwan1.html, available on 5th July 2019

\bibitem{...}
Technical documentation for evaluation board based on the SPIRIT1 https://www.st.com/en/evaluation-tools/steval-ikr002v1d.html, available on 5th July 2019

\bibitem{...}
Technical documentation for Semtech SX1276 137MHz to 1020MHz Long Range Low Power Transceiver https://www.semtech.com/products/wireless-rf/lora-transceivers/sx1276, available on 5th July 2019 

\bibitem{...}
CosmicWatch web page http://www.cosmicwatch.lns.mit.edu, available on 5th July 2019

\bibitem{...}
Technical documentation for STM32F446xC/E microcontroller https://www.st.com/resource/en/datasheet/stm32f446ze.pdf, available on 5th July 2019 

\bibitem{...}
Technical documentation for FGPMMOPA6H GPS Standalone Module https://cdn-shop.adafruit.com/datasheets/GlobalTop-FGPMMOPA6H-Datasheet-V0A.pdf, available on 5th July 2019 




\end{thebibliography}
\end{document}